\documentclass[twocolumn,showpacs,preprintnumbers,amsmath,amssymb, prb]{revtex4}
\usepackage{graphicx}
\usepackage{dcolumn}
\usepackage{bm}
\usepackage{wasysym}
\usepackage{textcomp}
\usepackage{epstopdf}
\usepackage{stmaryrd}
\usepackage{txfonts}

\begin{document}

\title{Exchange bias-like magnetic properties in Sr$_{2}$LuRuO$_{6}$}

\author{R.\,P. Singh and C.\,V. Tomy}

\affiliation{Department of Physics, Indian Institute of Technology Bombay, Mumbai-400076,
India.}

\date{\today}

\begin{abstract}
Exchange bias properties are observed in a double perovskite compound,
Sr$_{2}$LuRuO$_{6}$. The observed exchange bias properties have
been analyzed on the basis of some of the available theoretical models.
Detailed magnetization measurements show that the exchange bias properties
are associated with the Dzyaloshinsky--Moria (D--M) interaction among
the antiferromagnetically ordered Ru moments ($T_{N}\sim32\,\mbox{K}$).
In addition to the usual canting of the antiferromagnetic moments,
D--M interaction in this compound also causes a magnetization reversal
at $T\sim26\,\mbox{K}$, which seems to trigger the exchange bias
properties. Heat capacity measurements confirm the two magnetic anomalies.
\end{abstract}

\maketitle

\section{Introduction}

Exchange bias (EB) usually refers to an offset in magnetization hysteresis
loop along the field axis \cite{Meiklejohn 1956} in an antiferromagnetic/ferromagnetic
system due to the unidirectional anisotropy when the magnetization
is measured after cooling down the sample in an external magnetic
field below its magnetic ordering temperature. This effect is usually
observed in some magnetic nanoparticles and thin films containing
antiferromagnetic/ferromagnetic or ferromagnetic/spin-glass bilayers.
However, this effect is also observed in some bulk materials such
as manganites, cobaltates, intermetallic compounds \cite{J. Nogus},
spin-glass systems, etc. Exchange bias is of immense technological
importance since it enables the control of reference magnetization
in spintronic devices such as read-heads and nonvolatile memory. From
a scientific point of view, it is of general interest because it involves
a sophisticated interplay between fundamental magnetic properties
such as anisotropy and exchange interaction as well as ferromagnetic
and antiferromagnetic order. Despite the intensive research, especially
in the last two decades, an understanding of the underlying microscopic
coupling mechanism is still missing.

Very recently, Dong et.\,al.\,\cite{Dong} has used Dzyaloshinsky--Moria
(D--M) interaction \cite{Dzyaloshinsky,Moriya } as the possible mechanism
for exchange bias in perovskites with compensated $G$-type antiferromagnetism.
There are not many compounds which exhibit D--M interaction and hence
detailed studies are not available regarding the EB like properties
exhibited by them. We have found that the perovskite compound, Sr$_{2}$LuRuO$_{6}$,
exhibits exchange bias-like properties from our magnetization measurements.
The D--M interaction generally occurs when the crystal structure has
low symmetry. The family of double perovskite compounds Sr$_{2}$$Ln$RuO$_{6}$
(where $Ln$ = Y or rare earth) form in a monoclinic structure (space
group $P2_{1}/n$) \cite{Battle 1984} and are good candidates for
exhibiting D--M interaction \cite{Greatrex,Doi } due to distortion
of the oxygen octahdra \cite{Ovchinnikov}. Since there are no reports
available for the detailed magnetic properties of Sr$_{2}$LuRuO$_{6}$,
we first give the detailed magnetic properties of this compound, which
is followed by the exchange bias studies. Heat capacity measurements
confirm the two anomalies, corresponding to the two magnetic orderings.
Our results further show that this compound shows exchange bias properties
associated with the D--M interaction.

\section{EXPERIMENTAL DETAILS}

Polycrystalline sample of Sr$_{2}$LuRuO$_{6}$ was prepared by the
standard solid state reaction method. Stoichiometric amounts of SrCO$_{3}$,
Lu$_{2}$O$_{3}$ and Ru metal powder were mixed thoroughly and heated
initially at $960^{\circ}$C for 24 hours. The samples were then given
two more intermediate heat treatments at $1350^{\circ}$C, before
pelletizing and sintering at $1360^{\circ}$C for 24 hours. Diffraction
pattern of the sample was recorded on an X'pert PRO x-ray diffractometer
(PANalytical, Holland). The Rietveld analyses of the x-ray diffraction
patterns using FULLPROF software showed that the compound forms in
a single phase. The observed patterns could be indexed to a monoclinic
structure with a space group $P2_{1}/n$. The lattice parameters obtained
from the analyses are $a$ = 5.727(2)\,$\textrm{\AA}$, $b$ = 5.727(2)\,$\textrm{\AA}$
and $c$ = 8.101(3)\,$\textrm{\AA}$ along with $\beta=90.2{}^{\circ}$,
which are in good agreement with those values reported earlier \cite{Battle 1989}.
The magnetization as a function of temperature and magnetic field
was measured using a vibrating sample magnetometer (VSM) attachment
of the Physical Property Measurement System (PPMS) (Quantum Design,
USA).

\section{RESULTS AND DISCUSSION}

\subsection{Magnetic properties}

\begin{figure}
\includegraphics[scale=0.42]{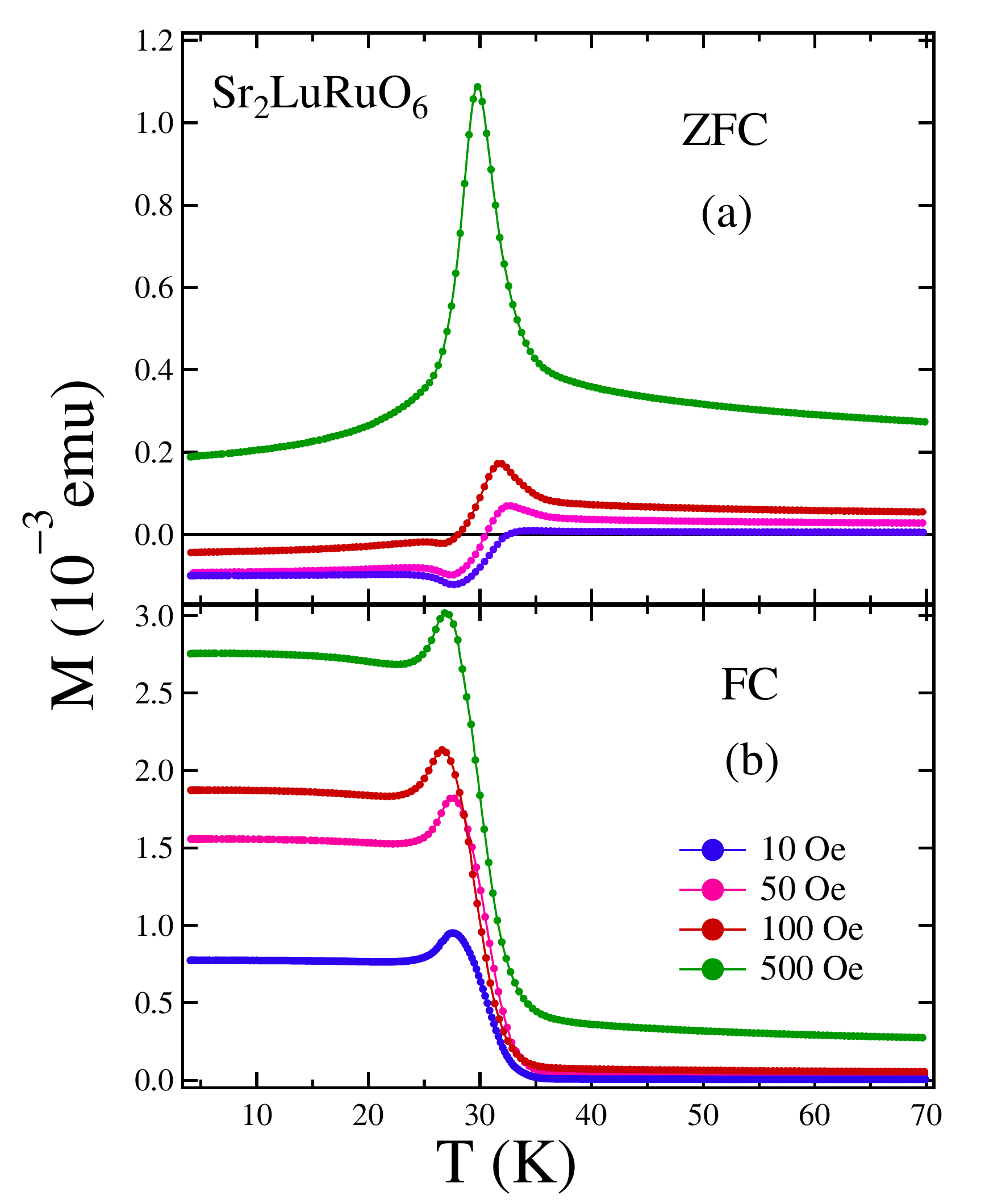}

\caption{Magnetization vs temperature for Sr$_{2}$LuRuO$_{6}$ in (a) zero
field-cooled (ZFC) and (b) field-cooled (FC) modes under various applied
fields.}

\end{figure}
Figure~1 illustrates the magnetization of Sr$_{2}$LuRuO$_{6}$ as
a function of temperature in zero field-cooled (ZFC) and field-cooled
(FC) modes. For low field values, the ZFC magnetization is negative
at lower temperatures. As the temperature is increased, the magnetization
decreases to go through a minimum, which is contrary to the normal
behaviour. As the temperature is further increased, the magnetization
increases, goes through a positive maximum, and then shows the normal
paramagnetic behaviour at high temperatures (not shown). For higher
fields ($\geq500$\,Oe), the magnetization is positive at all temperatures.
As shown in Fig.~1(b), the FC magnetisation is positive at all temperatures,
shows a broad peak centred around $T\sim26$\,K and the temperature
at which the peak occurs shows a weak temperature dependence on the
applied fields. The nature of increase in FC magnetization near $T_{N}\sim32$\,K
clearly indicates the presence of a ferromagnetic component which
is expected in this compound due to D--M interaction among the antiferromagnetically
ordered Ru moments. In order to confirm the presence of ferromagnetic
component as well as to find the origin of the anomalous behaviour
in magnetization, we have measured hysteresis loops at different temperatures.
Typical hysteresis loops are shown in Fig.~2(a)--(h) for selected
temperatures. At low temperatures, the hysteresis is negligible and
the loops are almost closed. Above $T\sim20$\,K, the hysteresis
loops open up and the coercivity increases as shown in Fig.~2(i).
Above $T\sim26$\,K, the hysteresis loops start closing again, the
coercivity decreases and disappears above the magnetic ordering temperature
of $T_{N}\sim32$\,K. As the sample is cooled through $T_{N}$, the
increase in coercivity can be attributed to the ferromagnetic component
developed due to the D--M interaction. The decrease in coercivity
below $T\sim26$\,K implies a net decrease in the ferromagnetic component.
The absence or negligible coercivity below $20$\,K can result from
the disappearance of the ferromagnetic component. However, this possibility
is quite unlikely since no structural change occurs in this compound
\cite{Battle 1989} and hence the D--M interaction (which gives rise
to canting) cannot vanish. Another possibility is the magnetization
reversal by which some of the Ru moments re-align opposite to the
ordered Ru moments reducing/cancelling the net ferromagnetic component.
This will explain the ZFC/FC magnetization behaviour also. When the
sample is cooled below $T_{N}$, the FM component increases the magnetization
in FC measurements. At $T=26$\,K, the reversal occurs, some moments
align opposite to the field and the magnetization decreases. Since
all the moments do not reverse/re-align, the magnetization does not
go to zero, but decreases and remains constant when the reversal is
complete. In ZFC measurements, even though the applied field is zero,
a small amount of negative remnant field can orient the moments along
the negative field direction at $T_{N}$ and some of the moments re-align
along the positive direction at the reversal temperature, resulting
in a mirror image of the FC curves. At the lowest temperature, when
the positive applied field is small, moments does not change the status
and hence the magnetization is negative. As the temperature is increased
to the magnetization reversal temperature, moments aligned along the
positive direction (those reversed), reverse back to the negative
direction, resulting in a sudden increase in the negative magnetization.
As the temperature is further increased, all the moments have to align
along the positive field direction when $T_{N}$ is approached, resulting
in a positive maximum.%
\begin{figure}
\includegraphics[scale=0.385]{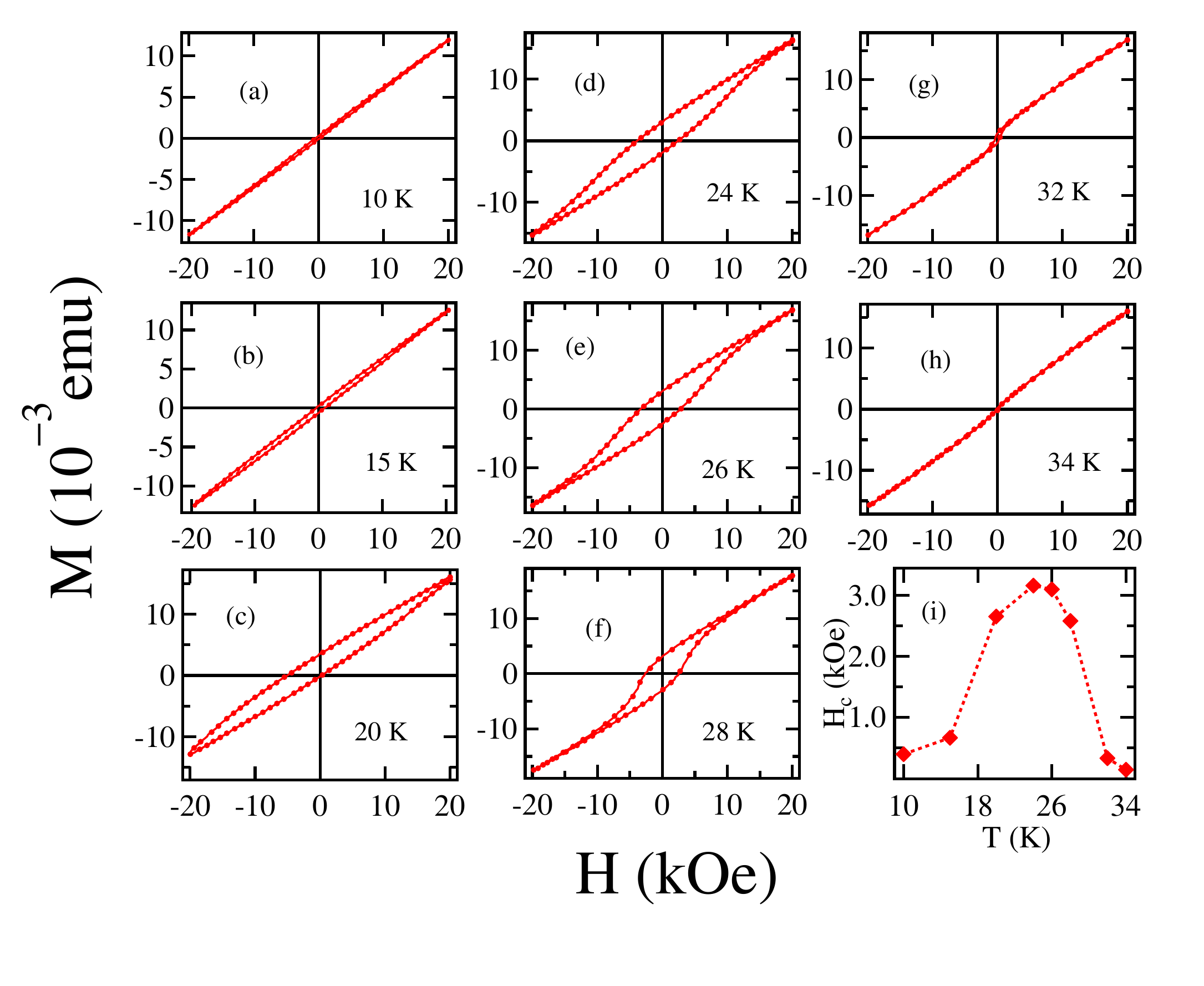}

\caption{Isothermal magnetization curves (a)--(h) obtained under ZFC mode for
Sr$_{2}$LuRuO$_{6}$ at different temperatures. (i) Coercivity ($H_{{\rm c}}$)
as a function of temperature obtained from the same curves.}

\end{figure}
\begin{figure}
\includegraphics[scale=0.42]{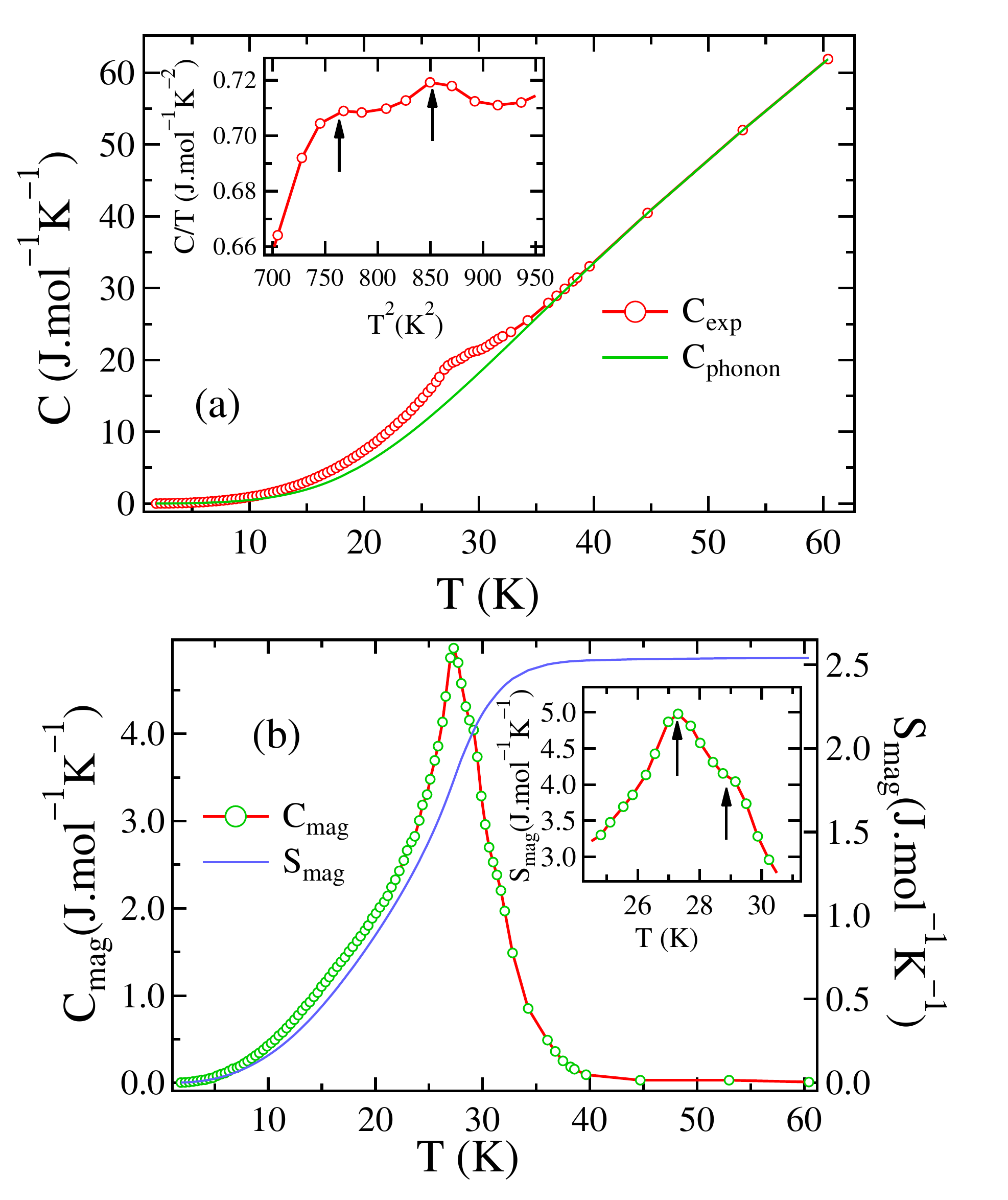}

\caption{Total measured heat capacity (circles) as a function of temperature.
The inset shows the expanded portion near the magnetic ordering temperatures.
Arrows indicate the two magnetic anomalies. Solid line is the calculated
phonon contribution. (b) left scale : magnetic contribution to heat
capacity ($C_{{\rm mag}}$) obtained by subtracting the calculated
phonon contribution from the measured heat capacity as a function
of temperature. Inset shows the expanded portion near the magnetic
anomalies. The magnetic entropy ($S_{{\rm mag}}$) is plotted on the
right scale.}

\end{figure}

Heat capacity measurements were carried out in the temperature range
2--200\,K. Figure~3(a) shows the measured heat capacity. Two anomalies
can be noted around the magnetic ordering temperatures (see the expanded
portion near the magnetic ordering in the inset of Fig.~3(a)). In
order to obtain the magnetic contribution to heat capacity, the phonon
contribution needs to be separated from the total heat capacity. As
there is no nonmagnetic analogue available for this compound, the
total heat capacity above the magnetic ordering temperature was fitted
to an equation containing standard Einstein and Debye terms \cite{Martin},

\begin{multline}
C_{{\rm ph}}=R\left(\frac{1}{1-\alpha_{D}}\left(\frac{\theta_{D}}{T}\right)^{3}\int_{0}^{x}\frac{x^{4}e^{x}}{\left(e^{x}-1\right)^{2}}dx+\sum_{i=1}^{3n-n}\frac{1}{1-\alpha_{E}}\,\frac{y^{2}e^{y}}{(e^{y}-1)^{2}}\right)\end{multline}
where $\alpha_{E}$ and $\alpha_{D}$ are the anharmonicity coefficients,$\theta_{D}$ is the Debye temperature, $\theta_{E}$ is the Einstein
temperature, $x=\theta_{D}/T$ and $y={\theta_{E}}_{i}/T$. The best
possible fit was obtained when the calculations were performed by
using one Debye and three Einstein frequencies along with a single
$\alpha_{E}$. The solid line in Fig.~3(a) represents the fit to
the phonon contribution (extended over the whole temperature range),
which is in good agreement with the experimental data at high temperatures
(above the magnetic ordering). The parameters obtained from the best
fit are: $\theta_{D}=228$\,K, $\theta_{E1}=130$\,K, $\theta_{E2}=507$\,K,
$\theta_{E3}=510$\,K, $\alpha_{E}=9.0\times10^{-4}$\,K$^{-1}$
and $\alpha_{D}=9.5\times10^{-5}$\,K$^{-1}$. The calculated phonon
contribution was then subtracted to obtain the magnetic heat capacity,
which is shown in Fig.~3(b). The well-defined peak indicates the
presence of a clear long range magnetic ordering. However, a closer
look can reveal a shoulder, indicated by the arrow, consistent with
two transitions in the magnetization measurements. The slight broadening
below the magnetic ordering may be caused by the subtraction of the
extrapolated phonon contribution, which was calculated only above
the magnetic ordering temperatures. The magnetic entropy calculated
from the magnetic heat capacity ($S_{{\rm mag}}\sim2.56$\,J\,mol$^{-1}$\,K$^{-1}$)
is also found to be well below the expected value even for a doublet
ground state of Ru ($S_{{\rm mag}}\sim5.76$\,J\,mol$^{-1}$\,K$^{-1}$),
consistent with the findings in other members of this family \cite{CVT PRB,CVT JPCM}.
Our preliminary results from inelastic neutron scattering measurements
\cite{INS} have indicated the presence of crystal field levels for
Ru in Sr$_{2}$YRuO$_{6}$. This crystalline field effect can reduce
the ground state of Ru$^{5+}$ to a doublet ground state. Further,
we have observed \cite{INS,Diff} the presence of diffuse scattering
even well above the magnetic ordering temperature. This diffuse scattering
is expected from the frustration among the Ru moments. If frustration
exists among the magnetic moments, it can reduce the entropy of spins
while entering the magnetically ordered state. If we consider similar
mechanisms operating in Sr$_{2}$LuRuO$_{6}$ also, then the decrease
in magnetic entropy in the present case can be attributed to the crystalline
electric field effects and frustration among the Ru moments.%
\begin{figure}[b]
\includegraphics[scale=0.5]{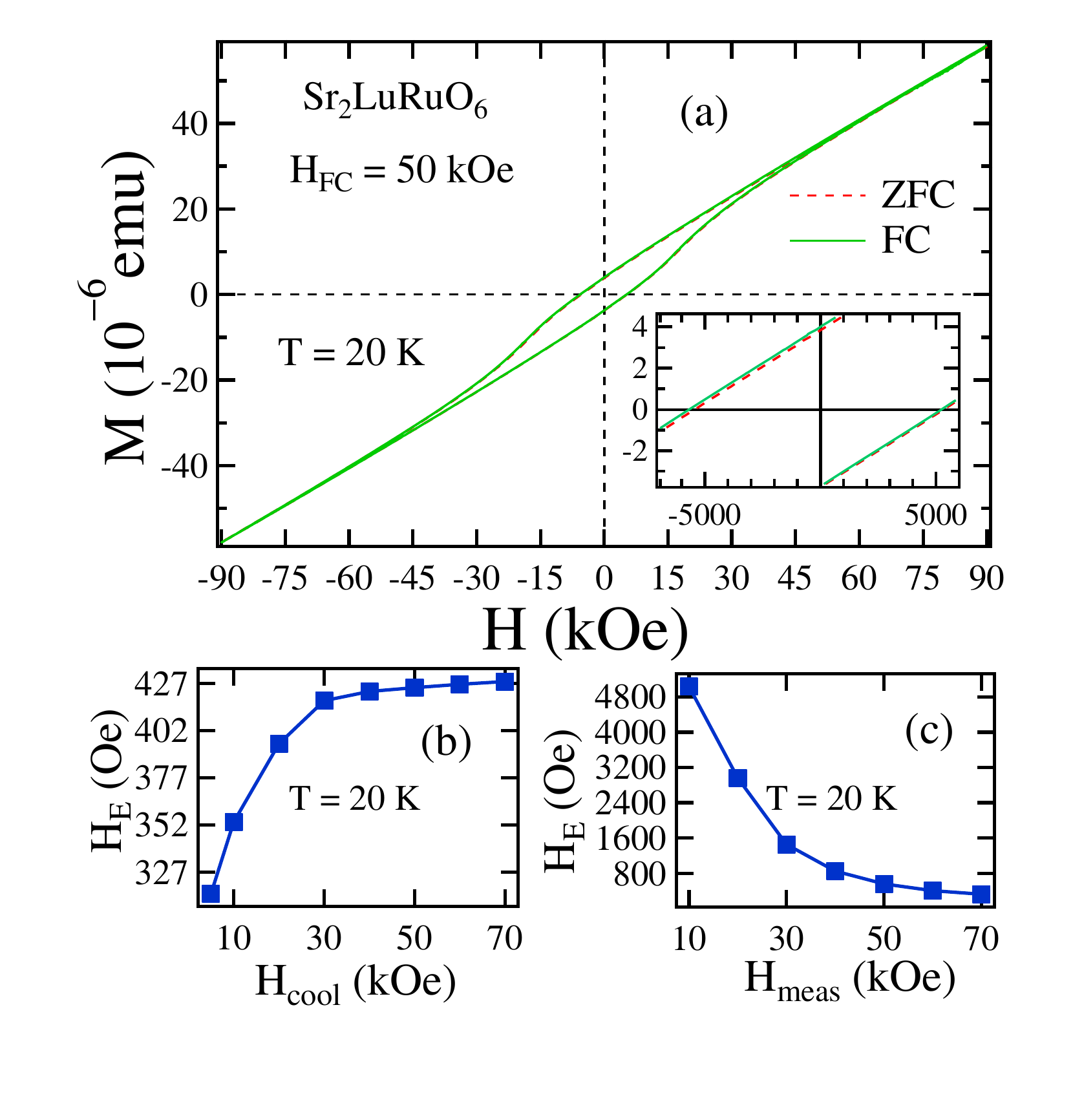}

\caption{(a) Expanded portion of magnetization hysteresis ($M$--$H$) loops
near origin for Sr$_{2}$LuRuO$_{6}$ at 20\,K measured in zero field-cooled
(ZFC) and field-cooled (FC) modes. For FC mode, the cooling field
($H_{{\rm FC}}$) is 50\,kOe . (b) Exchange bias (EB) field ($H_{E}$)
values deduced from FC hysteresis loops as a function of measuring
field, $H_{{\rm meas}}$. (c) Cooling field dependence of $H_{E}$
where the EB field values are plotted as a function of cooling field
($H_{{\rm FC}}$).}

\end{figure}

\subsection{Exchange bias-like properties}

In order to further verify the detailed magnetic behaviour, we have
measured the exchange bias properties of this compound. One of the
methods to verify the presence of exchange bias is to measure the
hysteresis loop after field-cooling (FC) the sample below its $T_{N}$,
and compare it with the hysteresis loop obtained in ZFC mode. For
the FC process, the sample was cooled to 20\,K in a magnetic field
of 50\,kOe (in zero applied field for ZFC process) and the hysteresis
loops were then measured between \textbf{$\pm$}\,90\,kOe, as shown
in Fig.~4(a). A clear difference between the two hysteresis loops
can be resolved (see the inset of Fig.~4(a) where the expanded portion
near the origin is plotted); while the ZFC hysteresis loop is centred
at zero field, the FC hysteresis loop shifts towards both the negative
field and the positive magnetization direction. This shift was further
confirmed by measuring the FC hysteresis in a cooling field of $-$50\,kOe,
which showed the shift along positive field and negative magnetization
direction. We define the shift along the field axis as the exchange-bias
field $H_{E}$ ($=-(H_{c+}+H_{c-})/2$) \cite{Niebieskikwiat }, where
$H_{c+}$ ($H_{c-}$) is the coercive field on the right (left) side
of the zero field. The observed value of $H_{E}$ is 216\,Oe, indicating
the presence of an exchange field in this compound. In EB measurements,
cooling field is known to affect the exchange bias. In order to select
a suitable cooling field, we measured the $M$--$H$ loops in various
cooling fields ranging from 5\,kOe to 70\,kOe at 20\,K. The exchange
field values calculated from the loop shift is shown in Fig.~4(b).
The exchange bias field shows a saturation type behaviour above 30\,kOe.
Another factor affecting the exchange bias measurement is the maximum
value of the field to which the field is cycled (minor hysteresis
loop effects). To circumvent this effect and select a suitable measuring
field, we measured hysteresis loops between different field values
after field-cooling the sample in a field of 50\,kOe. Loops obtained
at low cycling fields show large exchange bias values (see Fig.~4(c))
due to the fact that the loops are not closed or not reached the saturation
(minor hysteresis). However, as the cycling field exceeds $\pm$\,30\,kOe,
the loops start closing and the exchange field tends to saturate.
Since we have used $H_{{\rm FC}}=50$\,kOe and a measuring field
of 60\,kOe, we can ascertain that the exchange field obtained in
our measurements is the inherent property of the sample and not an
artefact of the measuring field or cooling field \cite{Geshev}.%
\begin{figure}
\includegraphics[scale=0.30]{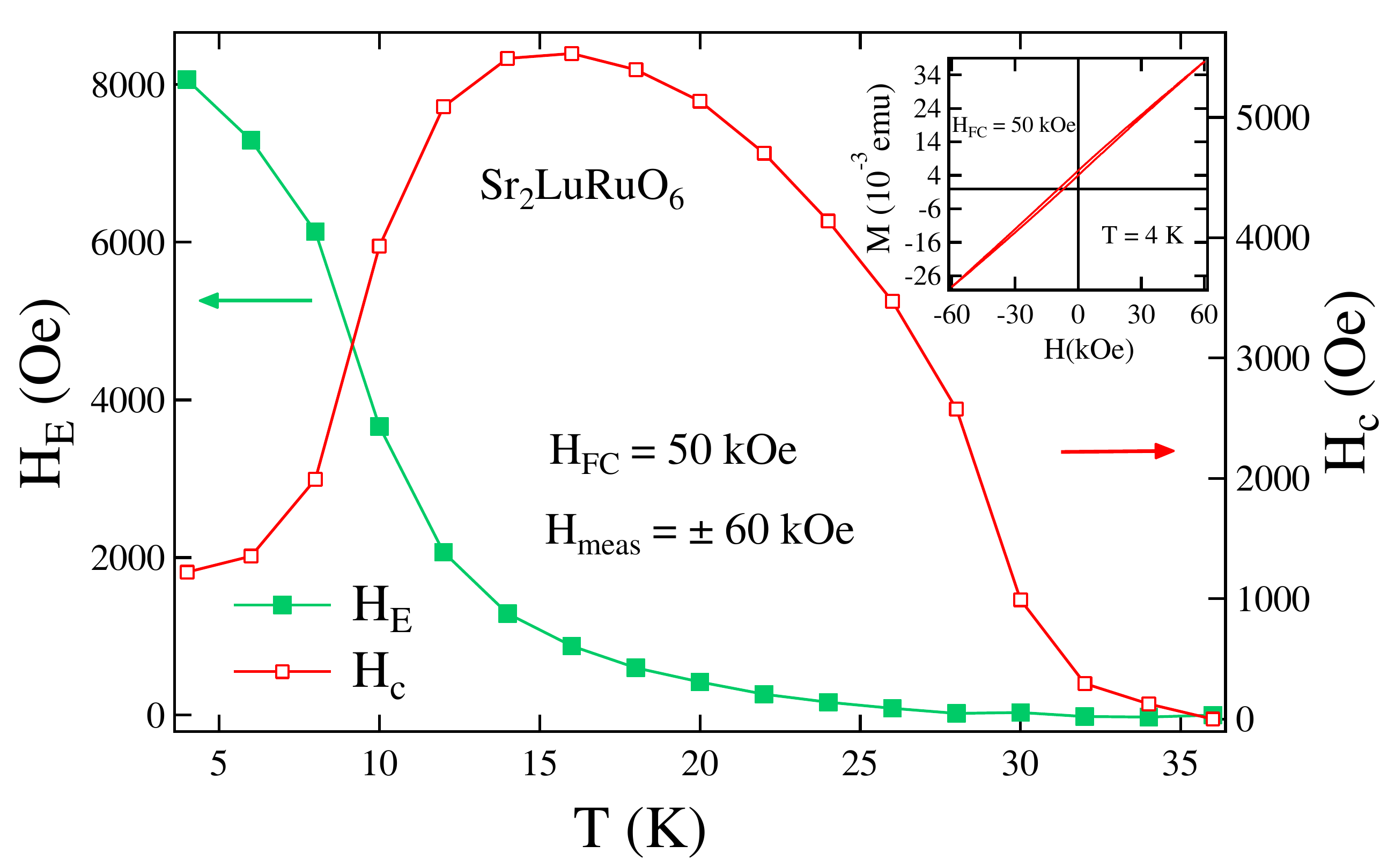}

\caption{(left scale): Temperature variation of the EB field $H_{{\rm E}}$.
The $H_{{\rm E}}$ values were deduced from the FC hysteresis loops
obtained at each temperature. (right scale): Temperature variation
of coercivity $H_{{\rm c}}$ obtained from the same hysteresis loops.
Inset shows a typical hysteresis loop at low temperatures (4\,K),
showing negligible coercivity, but large offset causing large $H_{{\rm E}}$
values.}

\end{figure}

In order to verify whether the observed EB like properties are associated
with the magnetic ordering, we studied the temperature dependence
of exchange bias. In these measurements, the sample was field-cooled
to the measuring temperature in an applied field of 50\,kOe. Once
the measuring temperature was reached, the magnetization loops were
measured between  $\pm$\,60\,kOe. Figure~5 shows variation of
the exchange field $H_{E}$ at different temperatures, along with
a plot of coercivity $H_{c}$. At 4\,K, the exchange field is as
high as 8000\,Oe. This happens due to a large offset of the hysteresis
loop along the magnetization/field axis (see inset of Fig.~5). As
the temperature is increased, $H_{E}$ value falls sharply and almost
disappears around 26\,K, even though coercivity and hence FM component
persists upto 32\,K, which is the magnetic ordering temperature for
this compound. The sudden increase of exchange field at low temperatures
indicates that the origin of exchange bias properties in this compound
is not purely from the appearance of the FM component (and hence D--M
interaction). It is more associated with the second magnetic anomaly
at $T\sim26\,\mbox{K}$, which reverses the magnetization.

We have measured another important property called the training effect
to ascertain the exchange bias in our compound, which is described
as the decrease of the exchange bias when the systems is cycled through
several successive hysteresis loops. Inset of Fig.~6 shows the expanded
portion of the low-field region in the negative field quadrant. The
arrow indicates the direction of increase in the field cycle ($n$).
The curves in the positive field quadrant do not show any visible
change. It is clear that the training effect is present in our sample
and the exchange bias decreases as the number of cycles is increased,
as shown in the main panel of Fig.~6, where the exchange bias field
is plotted as a function of $n$, the number of times through which
the sample is cycled in the field. The dependence of exchange-bias
on the number of field cycles ($n)$ is usually given by a simple
power-law relationship (for $n>1$) \cite{J. Nogus}.

\begin{equation}
H_{E}(n)-H_{E\infty}=k/\sqrt{n}\label{eq:power-law}\end{equation}
Here $H_{E}(n)$ is the exchange field at the $n^{th}$ cycle, $H_{E\infty}$
is the exchange field after a large number of field cycling ($n\rightarrow\infty$)
and $k$ is a system dependent constant. The solid line in Fig.~6
represents the best fit with the above empirical relation. We have
obtained $H_{E\infty}$ = 170\,Oe, which will be the remnant exchange
bias field in the sample. Since this power-law breaks down for $n=1$,
another approach to explain training effect is given by Binek \cite{Binek },
where the training effect in FM/AFM heterostructures is analyzed in
the framework of nonequilibrium thermodynamics. It was proposed that
the FM top layer, when consecutively cycled through magnetic fields,
triggers a spin configurational relaxation of the AFM interface magnetization
toward equilibrium. With this consideration, a recursive formula,
instead of the power-law formula, is given to describe the field cycling
dependence of exchange bias as,%
\begin{figure}
\includegraphics[scale=0.30]{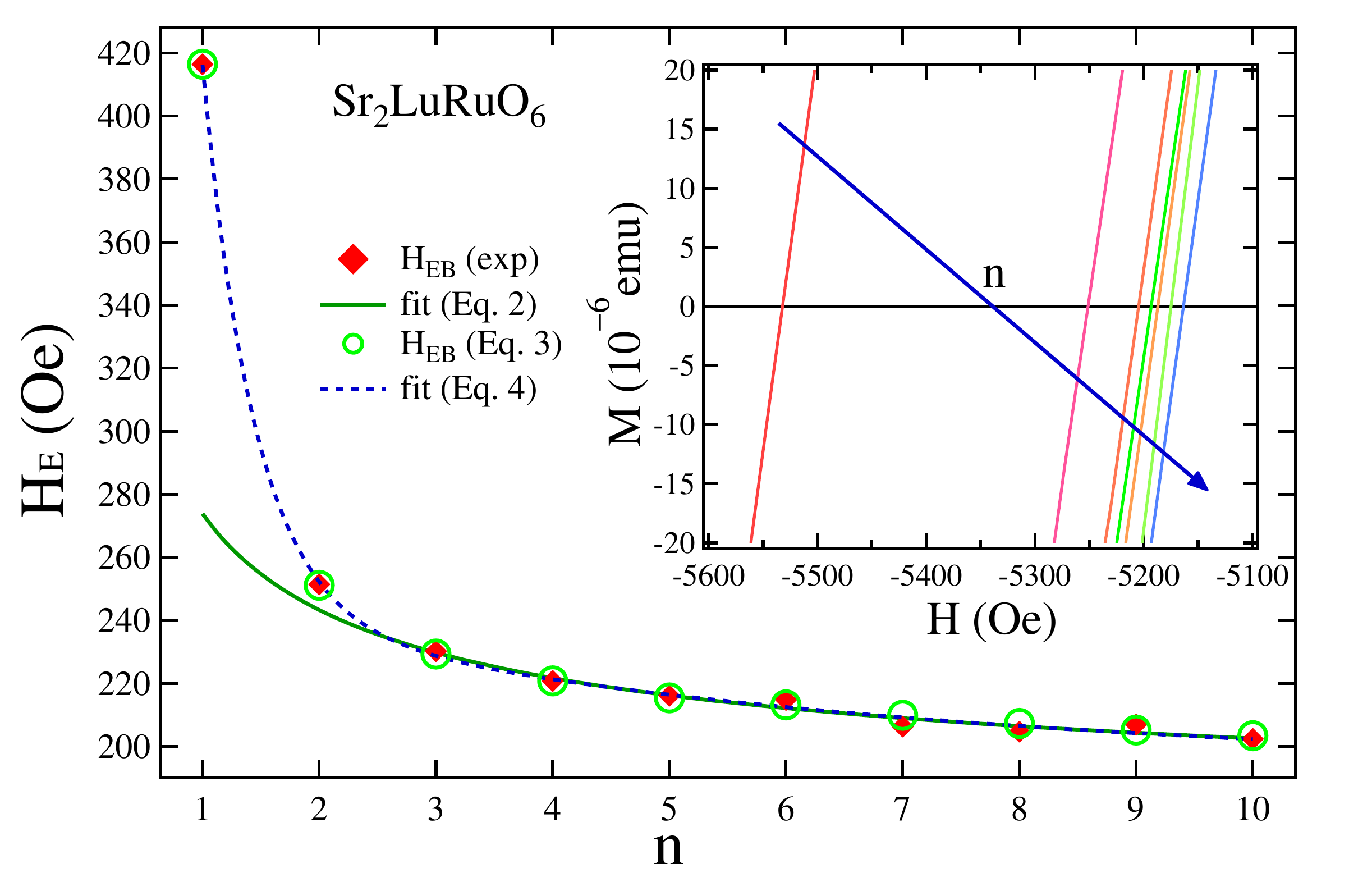}

\caption{The EB field ($H_{{\rm E}}$) dependence on the number of field cycles
($n$) (red diamonds) obtained from the training effect hysteresis
loops. The solid line represents the best fit for $H_{{\rm E}}$ data
to Eq.\,(\ref{eq:power-law}) for $n>1$. Open circles (green) are
the data points generated using the recursive formula, Eq.\,(\ref{eq:Binek eqn.}).
The dotted blue line shows the best fit with Eq.\,(\ref{mishra}).
Inset shows part of the hysteresis loops, demonstrating the training
effect due to exchange bias in Sr$_{2}$LuRuO$_{6}$. Arrow indicates
the direction of increase in field cycle ($n$).}

\end{figure}
 \begin{equation}
H_{E}(n+1)-H_{E}(n)=-\,\gamma\left(H_{E}(n)-H_{E\infty}\right)^{3}\label{eq:Binek eqn.}\end{equation}
where $\gamma$ is again a sample dependent constant ($=1/2k^{2}$).
We have attempted to generate data with Eq.\,(\ref{eq:Binek eqn.})
using $\gamma=4.1\times10^{-5}$\,Oe$^{-2}$ and $H_{E\infty}$ =
170\,Oe (obtained from the power law fit, Eq.\,(\ref{eq:power-law}))
as additional inputs. The open (green) circles in Fig.~6 represent
the $H_{E}$ values generated by Eq.\,(\ref{eq:Binek eqn.}), which
match satisfactorily with the experimental data. Thus the spin configurational
relaxation model can describe our experimental results as well, which
prima-facie is not a multi-layer compound. It is likely that the consecutive
reversing of the FM component (generated by D--M interaction) triggers
the configurational relaxation of the interfacial AF spins toward
equilibrium and causes the training effect.

Recently Mishra et.\,al.~\cite{Mishra} considered the training
effect as related to the interfacial spin disorder. The evolution
of the interfacial disorder during the measurements of the hysteresis
loops causes a decrease of the exchange bias field. AF domain dynamics
is also found to affect the magnitude of coercive as well as exchange
bias fields. The combined effect of these contributions causes a gradual
decrease of exchange bias as a function of $n$. The exchange bias
then can be given by a probabilistic equation,\begin{equation}
H_{E}(n)=H_{E\infty}+A_{f\,}e^{-n/P_{f}}+A_{i}\, e^{-n/P_{i}}\label{mishra}\end{equation}
where $A_{f}$ and $P_{f}$ are parameters related to the change of
the frozen spins, $A_{i}$ and $P_{i}$ are parameters related to
the evolution of the interfacial disorder. The $A$ parameters have
dimension of field while the $P$ parameters have no dimension but
they are similar to a relaxation time, where the continuous variable
{}``time\textquotedblright{} is replaced by a discrete variable $n$.
We have used this equation to fit our experimental data, which is
shown as the dotted (blue) line in Fig.~6, which agrees well with
the experimental data. The parameters obtained from this fit are:
$H_{E\infty}=192$\,Oe, $A_{f}=55.8$\,Oe, $P_{f}=5.5$, $A_{i}=1628$\,Oe,
$P_{i}=0.45$. The value of $H_{E\infty}=192$\,Oe compares well
with the value ($H_{E\infty}$ = 170\,Oe) obtained from the power-law
fit (Eq.\,(\ref{eq:power-law})). A comparison between $P_{f}$ and
$P_{i}$ shows that the frozen component relaxes nearly 10 times faster
than the interfacial magnetic frustration.

\section{Conclusion}

We have shown that the perovskite compound, Sr$_{2}$LuRuO$_{6}$,
exhibits properties associated with exchange bias. The observed exchange
bias properties could be analyzed on the basis of some of the available
theoretical models. The D--M interaction can be considered as the
possible mechanism behind the anomalous magnetic behaviour shown by
this compound. However, the indications are that the exchange bias
properties become dominant only after the magnetization reversal,
and not along with the magnetic order with FM component. Detailed
measurements are needed to pinpoint the exact mechanism for the complex
magnetic behaviour in this compound. In short, out results support
the proposal by Dong et.\,al. that the perovskites with compensated
AFM structure and weak ferromagnetic component (coming from D--M interactions)
can show the exchange bias like properties.

\end{document}